\documentclass[a4paper]{article}
\usepackage{aqis}
\usepackage{graphicx}
\usepackage{subfigure}

\begin{document}

\title{
   A Measurement-Based Form of the Out-of-Place Quantum Carry-Lookahead Adder
}

\author{
  Agung Trisetyarso 
  \affiliation{1}
    \email{trisetyarso07@a8.keio.jp}
  \and
  Rodney Van Meter 
  \affiliation{2}
  \email{rdv@sfc.wide.ad.jp}
  \and
	Kohei M. Itoh 
  \affiliation{1}
  \email{kitoh@appi.keio.ac.jp}
}

\address{1}{
  Department of Applied Physics and Physico-Informatics, Keio University, \\
 Yagami Campus, 3-14-1 Hiyoshi, Kohoku-ku, Yokohama-shi, Kanagawa-ken 223-8522, Japan
}
\address{2}{
  Faculty of Environment and Information Studies, Keio University, \\ 
  Shonan Fujisawa Campus,	5322 Endo, Fujisawa-shi, Kanagawa 252-8520, Japan
}

\abstract{
We present the design of a quantum carry-lookahead adder using measurement-based quantum computation. The quantum carry-lookahead adder (QCLA) is faster than a quantum ripple-carry adder; QCLA has logarithmic depth while ripple adders have linear depth. Our design is evaluated in terms of number of time steps and the total number of qubits used.
}


\maketitle


Measurement-based quantum computation (MBQC) is a new paradigm for implementing quantum algorithms using a quantum cluster state \cite{Raussendorf01}\cite{Briegel01}\cite{Raussendorf03}. MBQC is attractive because cluster states are considered to be easy to create on systems ranging from the polarization state of photons \cite{Walther05} to charge qubits. Quantum information propagation in a cluster is driven by the pattern of measurement bases, regardless of the measurement outcomes \cite{Raussendorf01}\cite{Briegel01}. A cluster state is in the form of \begin{equation}\left|\Phi_{\textit{N}}\right\rangle =\frac{1}{2^{\textit{N}}}\otimes^{\textit{N}}_{\textit{a}=1}\left(\left|0\right\rangle_{\textit{a}}\sigma^{\textit{a}+1}_{\upsilon}+\left|1\right\rangle_{\textit{a}}\right)\end{equation}, where $\upsilon$ can be $\textit{x,y,}$ or $\textit{z}$ depending on the choice of interaction Hamiltonian between neighbors$^{\cite{Walther05}}$ and with the convention $\sigma^{\textit{N}+1}_{\upsilon}=1$. In general, the cluster state should obey the quantum correlation equation\begin{equation}\sigma^{\textit{a}}_{\textit{i}}\otimes\sigma^{\textit{b}}_{\textit{j}}\left|\Phi_{\left\{\textit{k}\right\}}\right\rangle_{\textit{C}}=\left(-1^{\textit{k}_{\textit{a}}}\right)\left|\Phi_{\left\{\textit{k}\right\}}\right\rangle_{\textit{C}}\end{equation}
\textit{i}$\neq$\textit{j}=\textit{0,x,y,z} and $\textit{k}\textit{$_{a}$}=\left\{0,1\right\}$ where the upper index $\left(\textit{a}\right)$ represents a cluster site in the lattice and $\left(\textit{b}\right)$ is its neighbor site. The binary parameters $\textit{k}\textit{$_{a}$}$ are a set of binary parameters specifying the cluster state.

We consider a two-dimensional rectangular lattice with Manhattan geometry. Employing quantum correlations for quantum computation, as stated in Raussendorf's first theorem in \cite{Raussendorf03}, quantum gates can be simulated by measuring lattice qubits in a particular basis. All gates in the Clifford group, including CNOT, can be performed in one time step via a large number of concurrent measurements. Remarkably, because both wires and SWAP gates are in the Clifford group, MBQC supports long-distance gates in a single time step.  The Toffoli Phase gate can be executed in two time steps, where the measurement basis for the second step is adapted depending on previous measurement outcomes.Those benefits can be seen as a potential resource to attack complex problems such as the quantum carry-lookahead adder.


\begin{figure*}
\caption{MBQC QCLA Circuit}
\begin{center}
\includegraphics[height=90mm, width=170mm]{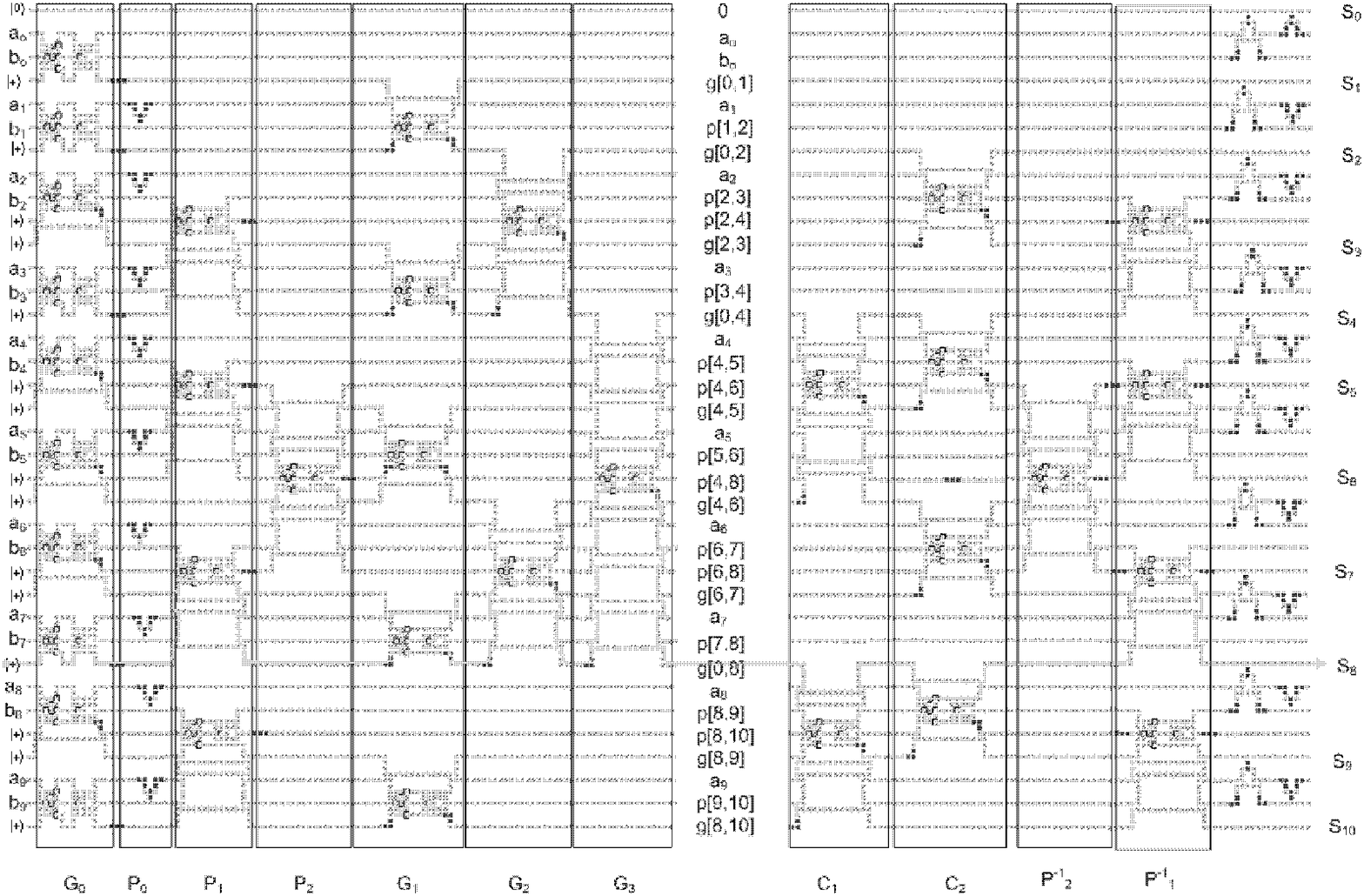}
\end{center}
\end{figure*}

Addition is a critical subroutine for algorithms such as Shor's algorithm for factoring large numbers \cite{Sho97SIComp}. Addition can be executed in many ways, with its performance being primarily dependent on carry propagation. The simplest method is ripple-carry addition, which has depth of O(\textit{n})\cite{Vedral96}. In a ripple-carry adder, carry information is propagated from the low-order qubits to the high order qubits one step at a time.  

Raussendorf et al. mapped the VBE ripple-carry adder to MBQC\cite{Vedral96}\cite{Raussendorf03}. However, a ripple-carry adder does not take good advantage of the strengths of MBQC. By unifying the Quantum Carry-Lookahead Adder (QCLA) with MBQC, we have designed a much faster circuit.

\begin{figure*}[!t]
\caption{Size and Depth Comparison between MBQC VBE and MBQC QCLA}
\begin{center}
\subfigure[MBQC VBE vs. MBQC QCLA  Depth] 
{
    \label{fig:sub:a}
    \includegraphics[height=45mm]{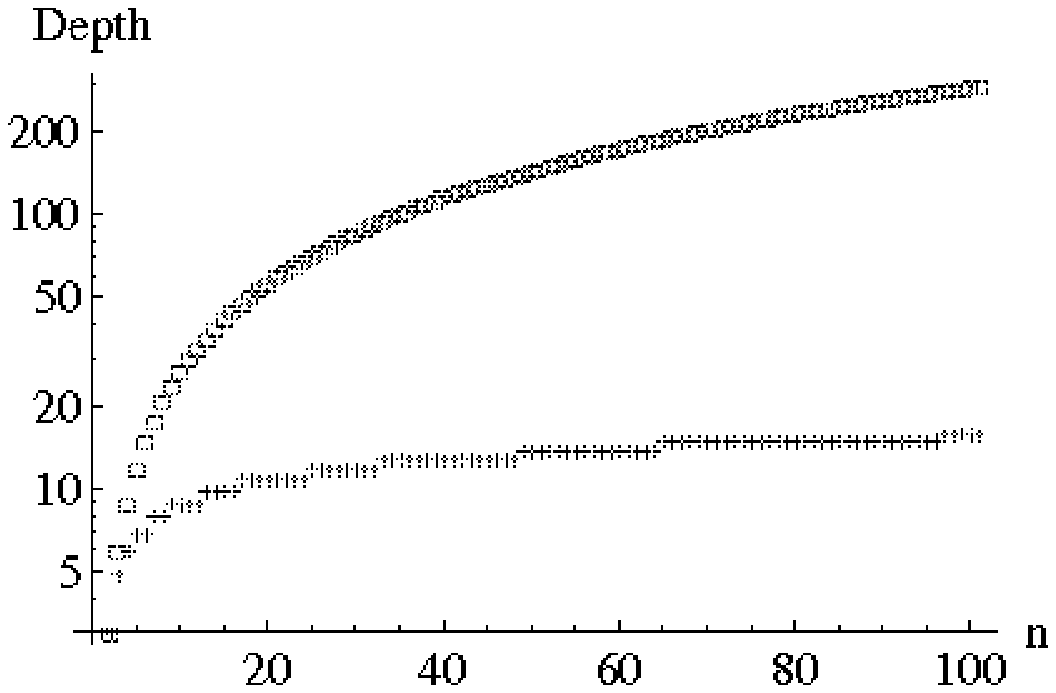}
}
\hfil
\subfigure[MBQC VBE vs. MBQC QCLA  Size] 
{
    \label{fig:sub:b}
    \includegraphics[height=45mm]{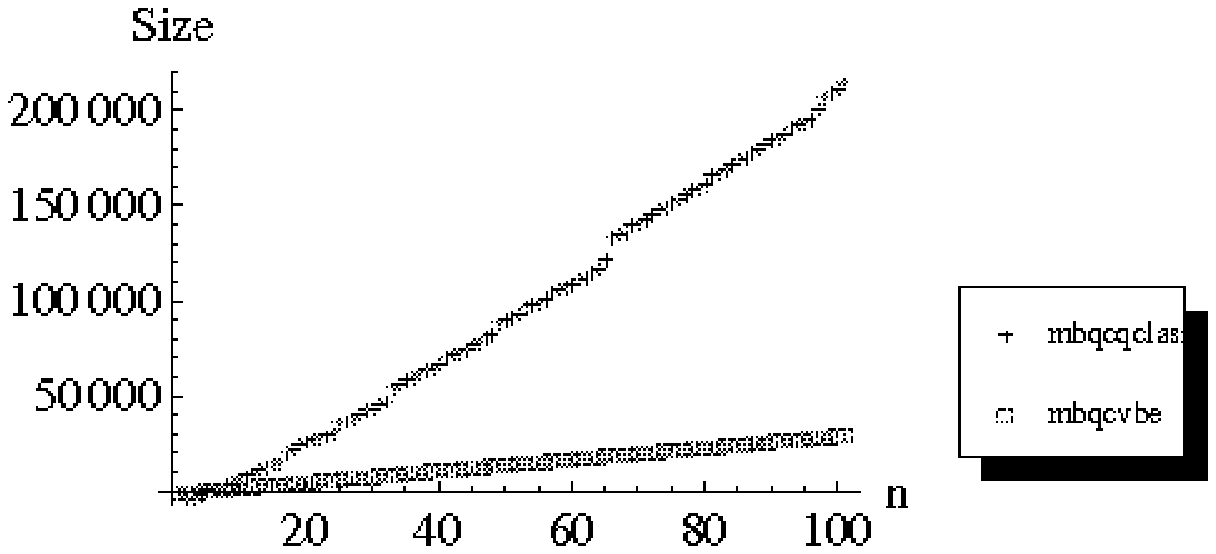}
}
\label{fig:sub} 
\end{center}
\end{figure*}

The quantum carry-lookahead adder is potentially more efficient than a quantum ripple-carry adder since its depth is O(log \textit{n}) \cite{Draper06}. A carry-lookahead adder uses three phases, the "Generate", "Propagate", and "Kill" networks, each of which progressively doubles the length of its span in each time step. In practice, the networks are somewhat redundant, and Draper et al. defined their circuit using only the P, C and G networks to calculate the final carry. Unfortunately, QCLA requires long-distance gates. The out-of-place form of the QCLA performs the unitary transformation $\left|\textit{a,b,0}\right\rangle \longrightarrow \left|\textit{a,b,a}+\textit{b}\right\rangle \longrightarrow$ where $\left|\textit{a}\right\rangle, \left|\textit{b}\right\rangle$ and $\left|\textit{a}+\textit{b}\right\rangle$ are \textit{n}-qubit registers. 

Our design for a 10-bit form of the out-of-place QCLA on MBQC is shown in Fig.1. The input qubits are on the left (top in the rotated figure) and output states are on the right. The propagation pattern of one qubit is highlighted in yellow. Our logical qubits are spaced with a pitch of four lattice sites. Each large box outlines one round in the P, G or C networks. The circuit is presented in unoptimized form for clarity. 

In our circuit, the depth is reduced to $\lfloor$log$_{2}$(\textit{n})$\rfloor$+$\lfloor$log$_{2}$(\textit{n}/3)$\rfloor$+5 compared to $\approx$O(\textit{n}) for the ripple-carry. However, this circuit costs more in physical resources, $\approx$1011\textit{n}+224$n$$\times$$\lfloor$log$_{2}$(\textit{n})$\rfloor$ compared to $\approx$304\textit{n} for the ripple-carry. The comparison of size and depth between MBQC VBE and MBQC QCLA are shown in Fig.2.  

\paragraph{Acknowledgments}

This work was supported in part by Grant-in-Aid for
Scientific Research by MEXT, Specially Promoted Research
No. 18001002 and in part by Special Coordination Funds for
Promoting Science and Technology.


\begin{thebibliography}{9}

\bibitem{Raussendorf01}
R.~Raussendorf and ~H.~J.~Briegel.
\newblock A One-Way Quantum Computer.
\newblock {\em Phys.\ Rev. Lett.}, 86, 5188 (2001).

\bibitem{Briegel01}
H.~J.~Briegel and R.~Raussendorf.
\newblock Persistent Entanglement in Arrays of Interacting Particles.
\newblock {\em Phys. Rev. Lett.} 86, 910 (2001).

\bibitem{Raussendorf03}
Robert Raussendorf, Daniel E.~Browne, and Hans J.~Briegel.
\newblock Measurement-based quantum computation on cluster states.
\newblock {\em Phys. Rev. A} 68, 022312 (2003).

\bibitem{Walther05}
P.~Walther, K.~J.~Resch, T.~Rudolph, E.~Schenck, H.~Weinfurter, V.~Vedral, M.~Aspelmeyer and A.~Zeilinger.
\newblock {\em Nature (London) 434},
\newblock 176 (2005).

\bibitem{Sho97SIComp}
P.~W.~Shor.
\newblock Polynomial-time algorithms for prime factorization
and discrete logarithms on a quantum computer.
\newblock {\em SIAM J. on Comp.}, 26(5):1484--1509, 1997.

\bibitem{Vedral96}
Vlatko Vedral, Adriano Barenco, and Artur Ekert.
\newblock Quantum networks for elementary arithmetic operations.
\newblock {\em Phys. Rev. A}.~54, 147 (1996).

\bibitem{Draper06}
T. Draper, S. Kutin, E. Rains, and K. Svore.
\newblock A Logarithmic-Depth Quantum Carry-Lookahead Adder.
\newblock {\em J. on. QIC}~6, 4--5, 351-369 (2006).

\end{thebibliography}
\end{document}